# Parabolic negative magnetoresistance in p-Ge/Ge$_{1-x}$Si$_x$ heterostructures

Yu.G.Arapov[1], G.I.Harus, O.A.Kuznetsov[*], V.N.Neverov, N.G.Shelushinina

*Institute of Metal Physics, Ural Branch of RAS, 620219, Ekaterinburg, GSP-170, Russia*
*\*Scientific-Research Physico-Technical Institute of State University, 603600, N. Novgorod, GSP-105, Russia*

Quantum corrections to the conductivity due to the weak localization (WL) and the disorder-modified electron-electron interaction (EEI) are investigated for the high-mobility multilayer p-Ge/Ge$_{1-x}$Si$_x$ heterostructures at $T = (0.1 \div 20.0)$K in magnetic field $B$ up to 1.5T. Negative magnetoresistance with logarithmic dependence on T and linear in $B^2$ is observed for $B \geq 0.1$T. Such a behavior is attributed to the interplay of the classical cyclotron motion and the EEI effect. The Hartree part of the interaction constant is estimated ($F_\sigma = 0.44$) and the WL and EEI contributions to the total quantum correction $\Delta\sigma$ at $B = 0$ are separated ($\Delta\sigma_{WL} \approx 0.3\Delta\sigma$; $\Delta\sigma_{EEI} \approx 0.7\Delta\sigma$).

## 1. Introduction

The diffusive nature of electron motion in disordered conductors results in quantum corrections to the effects with nontrivial dependencies on temperature $T$ and magnetic field $B$ [1,2]. These corrections are of the order of $(k_F l)^{-1}$, where $k_F$ is the Fermi quasimomentum and $l$ is the impurity scattering length. The total quantum correction to the Drude conductivity consists of the single-particle weak localization part and the part due to the disorder-modified electron-electron (e-e) interaction between particles with close momenta and energies (in the diffusion channel) and between particles with small total momentum (in the Cooper channel). For two-dimensional (2D) system all three quantum corrections, i.e., localization, e-e interaction in the diffusion channel and e-e interaction in the Cooper channel lead to the logarithmic low-temperature dependence of the conductivity at $B = 0$:

$$\Delta\sigma(T) = \frac{e^2}{2\pi^2\hbar}\left\{\left[p + \left(1 - \frac{3}{4}F\right) - (p-1)\beta(T)\right] \times \ln\frac{kT\tau}{\hbar} - \ln\frac{\ln kT_c\tau/\hbar}{\ln T_c/T}\right\}. \quad (1)$$

The first term in square brackets of Eq. 1 associated with the weak localization. The second term is a quantum correction due to the e-e interactions (EEI) in the diffusion channel. The third term is the Maki-Thomson correction. The second term in figure brackets is a quantum corrections due to EEI in the Cooper channel.

The different quantum corrections may be separated by application of an external magnetic field as each quantum effect has its own range of characteristic magnetic fields [3]. In the absence of spin scattering the magnetoresistance associated with the weak localization is negative. For this effect there exist two characteristic fields: the field $B_\varphi$ of crossover from parabolic to logarithmic $B$ - dependence of magnetoresistivity ($B_j = \hbar c/4eL_j^2$, $L_j$ – being the inelastic scattering length) and the field $B_{tr} = \hbar c/2el^2$, where the magnetic length become less than the elastic scattering length. For the effect in Cooper channel the characteristic field $B_{int}$ is the field, where the magnetic length become less than the coherence length $L_T$.

The localization effect is totally suppressed for field $B >> B_{tr} = \hbar c/2el^2$ where the magnetic length becomes less than the elastic mean free path $l$ [4]. In this range, the only quantum correction to the conductivity is from EEI in the diffusion channel. In contrast to the $B$ sensitivity of WL effect the calculation for the EEI in the absence of spin effects [5 - 7] demonstrates that

$$\Delta\sigma_{xx} \equiv \Delta\sigma^{ee} = (e^2/2\pi^2\hbar)g\ln(kT\tau/\hbar), \Delta\sigma_{xy} = 0, \quad (2)$$

irrespective of the strength of the applied magnetic field. Here $\tau$ is the elastic relaxation time and the interaction constant $g = (1 - F_\sigma)$, where the first universal term is due to the exchange (Fock) part and the second ($F_\sigma$) is related to the direct (Hartree) part of the Coulomb repulsion.

By inverting the conductivity tensor [8] in the presence of EEI corrections we have for the magnetoresistivity,

$$\rho_{xx}^{ee}(B) = 1/\sigma_0 + (1 - (\omega_c\tau)^2)\Delta\sigma^{ee}/\sigma_0^2, \quad (3)$$

where $\sigma_0$ is the Drude conductivity and $\omega_c$ is the cyclotron frequency. The consequence of Eq. (3) is twofold: irrespective of temperature $\rho_{xx}(B) = 1/\sigma_0$ for $\omega_c\tau = 1$ and the interplay of the classical cyclotron motion and the EEI effect in the diffusion channel leads to the parabolic negative magnetoresistance with logarithmic temperature dependence:

$$\rho_{xx}^{ee}(B) - \rho^{ee}(0) = -(\omega_c\tau)^2\Delta\sigma^{ee}/\sigma_0^2 \approx -B^2\ln T.$$

## 2. Experimental results and discussion

We have investigated the conductivity and magnetoresistance of strained multilayer p-Ge/Ge$_{1-x}$Si$_x$ (x = 0.03) heterostructures with the hole densities $p = (2.4 \div 2.6)\cdot 10^{11}$cm$^{-2}$ and mobilities $\mu = (1.0 \div 1.7)\cdot 10^4$cm$^2$/Vs ($k_F l \geq 10$) on Ge layers at $T \geq 0.1$K in magnetic fields up to 1.5T. The conductivity at $B = 0$ varies as the logarithm of T in a wide temperature range ($0.1 \div 20.0$)K (Fig.1). For B perpendicular to Ge layers the negative magnetoresistance is observed in a whole range of magnetic fields up to $\omega_c\tau = 1$ (Fig.2). Due to a high mobility of holes only a small magnetic field $B_{tr} = 0.03$T is needed to suppress the effect of weak localization. The logarithmic dependence of $\Delta\sigma$ on T at $B \gg B_{tr}$ (Fig.3) unambiguously is the evidence of the EEI quantum corrections. Fig.4 demonstrates that at $B \geq 3B_{tr}$ the magnetoresistance is really parabolic. The intersection point of curves for different $T$ at $\omega_c \tau \cong 1$ is also really observed (see Fig.2).

---

[1] E-mail: arapov@imp.uran.ru

The extrapolation of $B^2$ dependencies to $B = 0$ according to Eq.(2) gives the values of $\rho^{ee}(0) = 1/\sigma_0 + \Delta\sigma^{ee}/\sigma_0^2$ for each T. From the universal value of $\rho_{xx}$ at $\omega_c \tau = 1$ we have the Drude conductivity $\sigma_0 = 12.4 e^2/h$. Then in accordance with Eq.(2) the Hartree interaction constant $F_\sigma = 0.44$ has been estimated. Due to the transparent parabolic $\rho_{xx}(B)$ dependence in a wide range of magnetic fields where the EEI contribution is dominant the separation of WL and EEI parts of the total quantum correction $\Delta\sigma$ at $B = 0$ becomes possible. The result for our structures is that $\Delta\sigma^{ee} \approx 0.7 \Delta\sigma$ and $\Delta\sigma_{WL} \approx 0.3\Delta\sigma$.

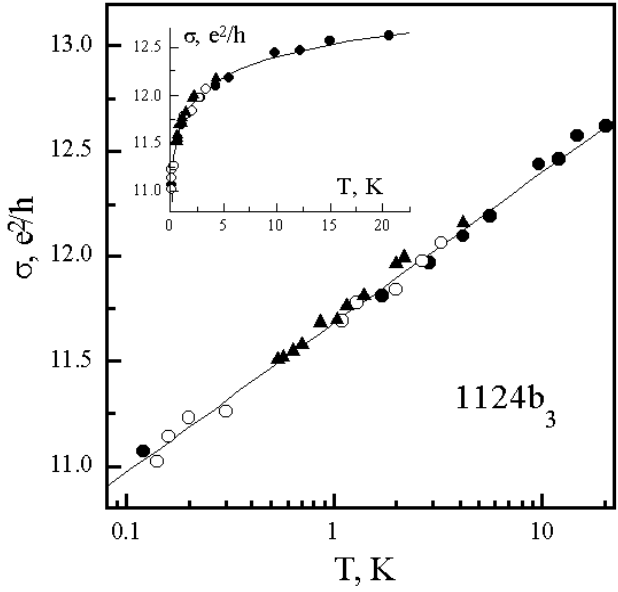

FIG. 1. The temperature dependence of the conductivity at $B$=0.

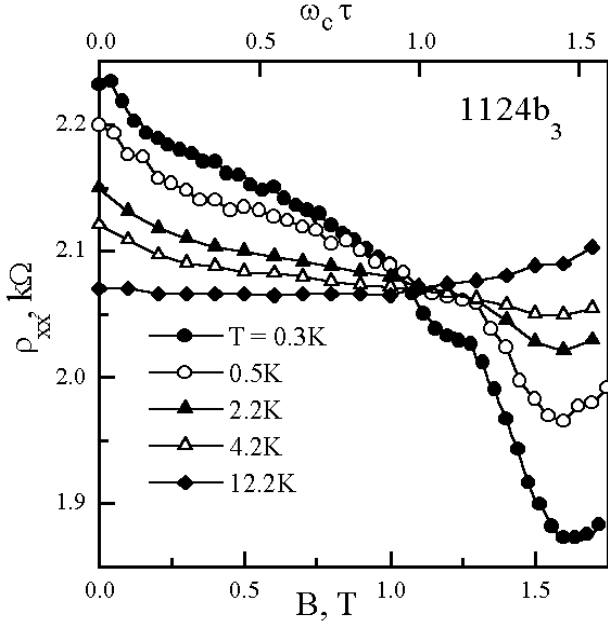

FIG. 2. The negative magnetoresistance as a function magnetic field at $T = 0.3 \div 12.2$ K.

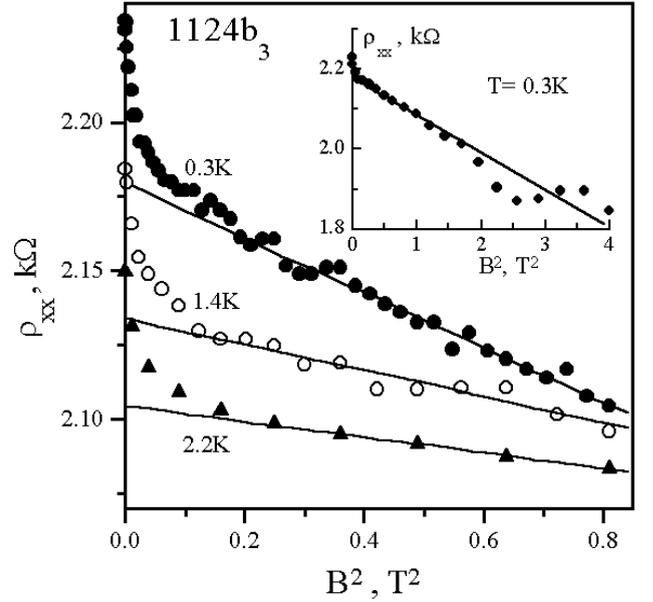

FIG. 3. Plot of the quantum correction to the conductivity $\Delta\sigma$ vs $\ln T$ for different magnetic fields.

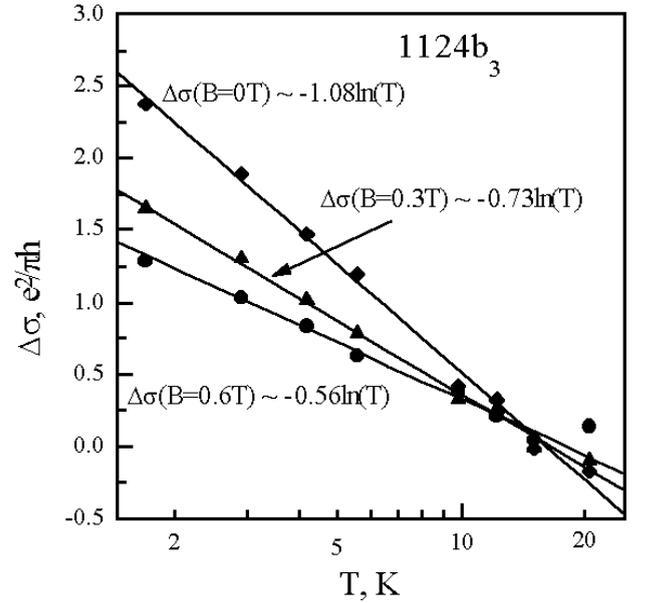

FIG. 4. The magnetoresistance $\rho_{xx}$ against $B^2$ from 0 to 0.9 T for $T$= 0.3 ÷ 2.2 K. The dashed line is the extrapolation of the $B^2$ dependence to zero field. The inset show one of curves (for $T$= 0.3K) at higher magnetic fields ($\omega_c\tau$ >1).

## 2. Conclusion

We have observed a large negative magnetoresistance of a high-mobility 2D-hole gas in p-$Ge_{1-x}Si_x$/Ge/$Ge_{1-x}Si_x$ quantum wells. We find that a negative magnetoresistance is proportional to $B^2$ and has a logarithmic temperature dependence. We attribute this behavior to the interplay of the classical cyclotron motion and the EEI corrections to the conductivity in the diffusion channel (exchange and Hartree contributions). Our sample parameters indicate that the weak localization and the EEI in the Cooper channel effects are totally suppressed in this field and temperature regime ($B_\varphi < 3 \cdot 10^{-4}$ T, $B_{int} < 0.03$ T and $T < 2.2$ K) but the Zeeman splitting is not get effective $B_s > 1$T above 2K. We find good agreement between theory and experiment oh the value of the Hartree interaction constant.

The work is supported by RFBR grant № 98-02-17306.